\documentclass[12pt]{article}
\usepackage{latexsym}
\usepackage{amsthm}
\usepackage{amsmath}
\usepackage[dvips]{graphicx}
\usepackage{amssymb}

\begin{document}
\baselineskip 0.6cm

\def\simgt{\mathrel{\lower2.5pt\vbox{\lineskip=0pt\baselineskip=0pt
           \hbox{$>$}\hbox{$\sim$}}}}
\def\simlt{\mathrel{\lower2.5pt\vbox{\lineskip=0pt\baselineskip=0pt
           \hbox{$<$}\hbox{$\sim$}}}}

\renewcommand{\l}{\langle}
\renewcommand{\r}{\rangle}
\newcommand{\be}{\begin{eqnarray}}
\newcommand{\ee}{\end{eqnarray}}

\newcommand{\dd}[2]{\frac{\partial #1}{\partial #2}}
\newcommand{\NN}{\mathcal{N}}
\newcommand{\LL}{\mathcal{L}}
\newcommand{\MM}{\mathcal{M}}
\newcommand{\ZZ}{\mathcal{Z}}
\newcommand{\WW}{\mathcal{W}}

\newcommand{\sslash}[1]{\ensuremath \raisebox{-.025cm}{$\diagup$}\hspace{-.425cm}#1\/}

\begin{titlepage}

\begin{flushright}
\end{flushright}

\vskip 2.5cm

\begin{center}

{\LARGE \bf 
Axion Protection from Flavor
}

\vskip 1.5cm

{\large Clifford Cheung$^{1,2}$}

\vskip 0.5cm

 {$^1$\it Department of Physics, University of California,
          Berkeley, CA 94720} \\
 {$^2$\it Theoretical Physics Group, Lawrence Berkeley National Laboratory,
          Berkeley, CA 94720} 
\vskip 0.5cm 
\texttt{clifford.cheung@berkeley.edu}

\vskip 1.5cm

\abstract

The QCD axion fails to solve the strong CP problem unless all explicit PQ violating, Planck-suppressed, dimension $n<10$ operators are forbidden or have exponentially small coefficients.  We show that all theories with a QCD axion contain an irreducible source of explicit PQ violation which is proportional to the determinant of the Yukawa interaction matrix of colored fermions.  Generically, this contribution is of low operator dimension and will drastically destabilize the axion potential, so its suppression is a necessary condition for solving the strong CP problem.  We propose a mechanism whereby the PQ symmetry is kept exact up to $n=12$ with the help of the very same flavor symmetries which generate the hierarchical quark masses and mixings of the SM.  This ``axion flavor protection'' is straightforwardly realized in theories which employ radiative fermion mass generation and grand unification.  A universal feature of this construction is that the heavy quark Yukawa couplings are generated at the PQ breaking scale.

\end{center}
\end{titlepage}

\section{Introduction}


The strong CP problem is elegantly solved by promoting the $\bar\theta$ parameter\footnote{Throughout this work, $\bar\theta$ will denote the physically observable strong CP phase, which includes the overall phase of the colored fermion mass matrix.} of QCD to a dynamical field known as the axion \cite{Peccei:1977hh,Peccei:1977ur,Weinberg:1977ma,Wilczek:1977pj}.  This is accomplished by introducing an anomalous PQ symmetry that is spontaneously broken, yielding a Nambu-Goldstone boson whose potential is generated non-perturbatively by QCD instantons.  When this axion dynamically relaxes to the minimum of its potential, the $\bar\theta$ parameter is effectively set to zero.

\begin{figure}[h]
\label{fig1}
\begin{center}
 \includegraphics[scale=1.25]{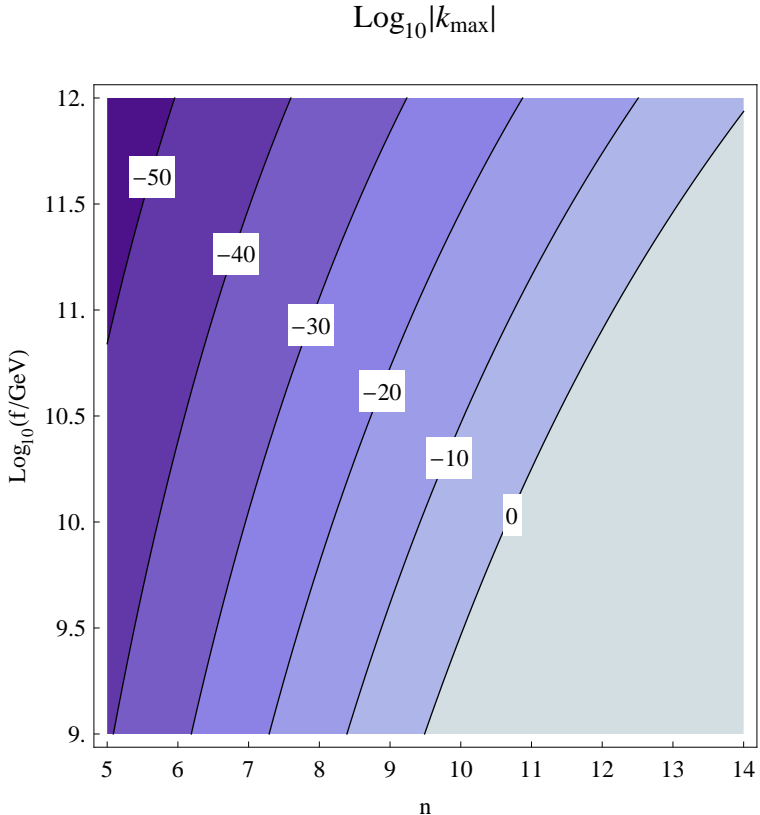} 
\end{center} 
\caption{A contour plot of $\log_{10} |k_{\rm max}|$, where $|k_{\rm max}|$ is the value of the operator coefficient $|k|$ at which the minimum of the axion potential is displaced by an amount greater than $\bar \theta = 10^{-10}$ for a given operator dimension $n$ and axion decay constant $f$.  At $k=k_{\rm max}$ the axion no longer solves the strong CP problem.  Here we have assumed that $\Lambda = m_{\rm Pl}$ and $\arg k_{\rm max} = \pi/4$.}
\end{figure}

In order for this mechanism to succeed, however, the axion must originate from a PQ symmetry which is of extraordinarily high quality \cite{Kamionkowski:1992mf,Kallosh:1995hi,Holman:1992us}.  In particular, if the PQ symmetry is spontaneously broken by a field $\phi$ at a scale $f$, then there will in general exist explicit PQ violating, dimension $n$ operators of the form
\be
{\cal O}_{\sslash{\textrm{PQ}}} &=& k \frac{\phi^n}{\Lambda^{n-4}} \qquad\overset{SSB}{\longrightarrow}\qquad |k|\frac{f^n}{\Lambda^{n-4}} \cos(na+\arg k),
\ee 
which can easily displace\footnote{It is conceivable that $\arg k = -n\bar\theta$, in which case the axion minimum is not displaced, but this would require an incredible fine-tuning.} the minimum of the axion potential by more than $\bar\theta =10^{-10}$ and effectively reintroduce the strong CP problem (see figure \ref{fig1}). In the most optimistic scenario, $\Lambda=m_{\rm Pl}$ is taken to be the Planck scale while $f=10^9$ GeV is taken to be as small as possible consistent with supernova constraints \cite{Turner:1989vc}.  Even so, if $|k|$ is of order unity then one requires $n\geq 10$ to successfully solve the strong CP problem.  Conversely, if the leading irrelevant operator, $n=5$, is to be adequately suppressed, then it is necessary that $|k| < 10^{-40}$.
Of course, the situation is even more dire if the axion decay constant is larger or if the fundamental gravity scale is low.

It has been argued that global symmetry violating operators of this kind should be induced at the Planck scale by quantum gravitational effects \cite{Bekenstein:1971hc,Bekenstein:1972ky,Giddings:1988cx,Coleman:1988tj,Gilbert:1989nq}.  For instance, a virtual black hole produced from some initial state of definite global charge will readily Hawking evaporate into a state of differing global charge---integrating out such processes yields Planck-suppressed, global symmetry violating operators at low energies.  

Because these results arise from quantum gravity, it is natural to consider string theoretic constructions in which non-perturbative violations of global symmetries are actually calculable.  In certain cases, one can identify PQ symmetries which are exact up to stringy instanton corrections of order $|k|\sim e^{-S}$, where $S$ is the string action evaluated on some background \cite{Svrcek:2006yi}.  This effectively reduces an extreme tuning to the logarithm of an extreme tuning.

While the stringy approach to PQ symmetry protection has its merits, 
it is important that we fully explore the limits of purely field theoretic alternatives.  This is the starting point of the present work.  In particular, we adopt the philosophy of the effective field theorist, which is that all allowed gauge invariant operators are naturally accompanied by order one coefficients.  Thus, any destabilizing contributions to the axion potential must be excluded for reasons of symmetry alone.  Along these lines, the state of the art in PQ symmetry protection has been to employ an ``automatic symmetry,'' i$.$e$.$ an accidental global symmetry which is exact up to some very high operator dimension as a consequence of a gauge symmetry \cite{Georgi:1981pu}.  According to this definition, baryon number is an automatic symmetry at the renormalizable level due to the SM gauge symmetry.  To this end, large discrete gauge symmetries such as $\mathbb{Z}_{n\geq 10}$ are conventionally invoked \cite{Krauss:1988zc,Alford:1990pt,Preskill:1990bm,Babu:2002ic,Dias:2002hz}. More recently, there have been some interesting alternative proposals for PQ symmetry protection which involve extra dimensions \cite{Cheng:2001ys} and anthropic considerations \cite{Carpenter:2009sw,Carpenter:2009zs}.


Conventional wisdom tells us that the stability of the axion potential is a question for the deep UV---in this paper we argue that this is not the case, and that the quality of the PQ symmetry is intimately connected to the dynamics which generates the masses of colored fermions.  We assert two principal claims, which are that
\newline
\begin{quote}
{ \it 1) All theories with a QCD axion contain an irreducible source of explicit PQ violation which can never be forbidden by non-anomalous gauge symmetries, and whose suppression is a necessary condition for solving the strong CP problem.}
\newline
\newline
{ \it 2) This and all other explicit PQ violating operators can be forbidden up to mass dimension $n=12$ with the help of the flavor symmetry of the SM.  This ``axion flavor protection'' simultaneously generates the observed hierarchy in the SM quark masses and mixings.
}
\newline 
\end{quote} 
To begin, consider point 1).  If $Q,\bar Q$ denotes the colored fermions and $\phi$ denotes the PQ symmetry breaking fields, then the interactions among these fields take the general form
\be
{\cal L} &\supset& Q \, M(\phi) \,\bar  Q,
\ee
where $M(\phi)$ is the Yukawa interaction matrix of colored fermions and all gauge and flavor indices have been suppressed.  
As we show in section \ref{sec:irred}, this implies the existence of an explicit PQ violating operator of the form 
\be {\cal O}_{\sslash{\textrm{PQ}}}&=&\det M(\phi),
\ee
 which is always allowed by gauge symmetries and may be induced by quantum gravitational effects. 
  This is a complete disaster for the axion because ${\cal O}_{\sslash{\textrm{PQ}}}$ must have operator dimension $n\geq 10$.  For instance, if $M(\phi)$ is simply linear in $\phi$, then $n$ is precisely equal to the number of colored fermions, which must in turn be extremely large.  Case in point, for the DFSZ axion we recognize this contribution as $\det M=h_u h_d$, i$.$e$.$ the $B\mu$ term, while for the KSVZ axion, $\det M=S$ where $S$ is a gauge singlet which gives a mass to the new heavy quarks.

This would all be very discouraging were it not for the SM quark mass hierarchy, which strongly suggests the existence of non-generic flavor structures which dictate a non-generic form for $M(\phi)$.  We are thus lead to point 2).  Indeed, the bulk of flavor-model building has been essentially devoted to swapping unnaturally small Yukawa couplings for higher dimension operators.  For example, in theories of Froggatt-Nielsen (FN) flavor \cite{Froggatt:1978nt,Babu:2009fd}, the mass ratios of the top, charm, and up quarks are given by
\be 
M_{33}:M_{22}:M_{11} &=& 1: \epsilon^2 : \epsilon^4,
\ee
where $\epsilon = \l\phi\r/m$ is set by the vev of some dynamical field $\phi$ and $m$ is the mass of some heavy fields which have been integrated out in order to generate the higher dimension operators. The story is similar for theories of radiative fermion mass generation \cite{Weinberg:1972ws,Georgi:1972mc,Georgi:1972hy,He:1989er,Hashimoto:2009xi,Babu:1990vx,Balakrishna:1987qd,Balakrishna:1988ks,Dobrescu:2008sz,Graham:2009gr,Babu:1989tv,ArkaniHamed:1996zw,ArkaniHamed:1995fq,Nandi:2008zw,Barr:2007ma,Barr:2008kg,Ibanez:1981nw} in which loop corrections generate the up mass from the charm mass and the charm mass from the top mass, and likewise for the quark mass mixings.  In either case, these constructions increase the overall operator dimension of $M(\phi)$, and can alleviate explicit PQ violation arising from $\det M(\phi)$.

In this paper, we construct a simple theory of radiative fermion mass generation in which explicit PQ violating operators are excluded up to very high operator dimension with the help of a gauged quark flavor symmetry.  Here the PQ symmetry is an automatic symmetry in part because of the SM flavor symmetry.  Moreover, the flavor structure simultaneously produces the correct hierarchy of masses and mixings for the SM fermions, and may be easily embedded in theories of grand unification.  This theory has the feature that the top Yukawa coupling is simultaneously generated by the PQ symmetry breaking dynamics.

Any field theory mechanism that successfully protects the axion from explicit PQ violating operators will necessarily increase the operator dimension of $\det M$.  However, many existing theories in the literature do so at the expense of flavor---that is to say, they invoke gauge symmetries which are ultimately inconsistent with the flavor hierarchies of the SM.  Thus, we find it compelling that one can construct simple theories which simultaneously stabilize axion potential and produce a natural theory of flavor.  


In section \ref{sec:irred}, we show that all axion theories contain an irreducible source of explicit PQ violation which is gauge invariant and, if generated, will reintroduce the strong CP problem.  Next, we present in section \ref{sec:model} a low energy description of all theories of radiative fermion mass generation and discuss the mechanics of axion flavor protection.  We conclude in section \ref{sec:conclusion}.   

\section{An Irreducible Source of Explicit PQ Violation}
\label{sec:irred}

Is there any choice of gauge group for which all explicit PQ violating, higher dimension operators are forbidden to all orders from the Lagrangian?  In this section we show that the answer to this question is no.  

Consider for the moment a theory of $N$ heavy quarks, $Q_i$ and $\bar Q^{\bar i}$, where the flavor indices $i$ and $\bar i$ each run from 1 to $N$ and color indices will be suppressed throughout.  Here $Q$ and $\bar Q$ transform in the fundamental and anti-fundamental representation of $SU(3)_c$, respectively. Furthermore, let $\phi$ denote all scalar fields or fermion bilinears which spontaneously break the PQ symmetry.  The interactions between $Q$, $\bar Q$ and $\phi$ take the general form
\be
{\cal L} &\supset& Q \, M(\phi) \,\bar  Q,
\ee
where $M(\phi)$ is the $N\times N$ interaction matrix of heavy quarks\footnote{If $SU(3)_c$ arises from GUT breaking, then this interaction will descend from operators contracting GUT matter multiplets with GUT breaking fields.} and matrix multiplication over quark flavor indices is implicit.  Note that $M(\phi)$ contains all interactions between the heavy quarks and the PQ breaking fields to {\it all} orders in higher dimension operators---thus, it takes the schematic form
\be 
M(\phi) &=& \phi + \frac{\phi^2}{m} + \frac{\phi^3}{m^2} +\ldots,
\ee
where $m$ is the scale of UV dynamics which has been integrated out.  Naturally, the heavy quark mass matrix, $M(\langle \phi \rangle)$, is generated only after PQ symmetry breaking.  For future convenience we rewrite the interaction matrix as
\be 
M(\phi) &=& m\, \lambda(\phi/m),
\ee
where $\lambda$ is an $N\times N$ matrix-valued polynomial.

In the absence of $\lambda$, there is an enhanced $G_{\rm flavor} = U(N)_L \times U(N)_{R}$ symmetry which acts on $Q$ and $\bar Q$.  Since all global and gauge symmetries necessarily reside within $G_{\rm flavor}$, we know that
\be
G_{\rm gauge} \times U(1)_{\rm PQ} \subseteq  G_{\rm flavor} ,
\ee
which means that under $G_{\rm gauge}$, the fields have to transform as
\be
Q &\rightarrow& QL^\dagger \\
\bar Q &\rightarrow& R \bar Q \\
\lambda &\rightarrow & L \lambda R^\dagger,
\ee
where $(L,R)\in G_{\rm flavor}$ and $(\det L )(\det R^\dagger) =1$ so that $G_{\rm gauge}$ does not have a color anomaly.  This immediately implies the existence of a gauge invariant operator
\be
{\cal O}_{\sslash{\textrm{PQ}}} &=& \det \lambda(\phi/m_{\rm Pl}),
\ee
which can always be added to the Lagrangian and will be induced by Planck scale dynamics.  Since there are no massless quarks, $\lambda$ has no zero eigenvalues and this operator cannot vanish.  Furthermore, the fact that $U(1)_{\rm PQ}$ is by definition anomalous with respect to $SU(3)_c$ implies that ${\cal O}_{\sslash{\textrm{PQ}}}$ is necessarily PQ covariant, so it is of course explicitly PQ violating!  This is obvious for the simplest possible PQ charge assignment, $\textrm{PQ}[Q]=\textrm{PQ}[\bar Q] =1$, which implies that $\textrm{PQ}[\det\lambda]=-2N$.

A similar story holds if we consider the SM quarks.  In this case, the interactions between the quarks and the PQ breaking fields are given by
\be
{\cal L} \supset q \, h_u \, \lambda_{u}(\phi/m) \,\bar u+q \,h_d \, \lambda_{d}(\phi/m) \,\bar d,
\ee
where $\lambda_{u,d}$ is a $3\times 3$ matrix-valued function and the electroweak gauge indices are contracted in the usual way.  As before, any new gauge symmetries must be contained in the quark flavor group $G_{\rm flavor} = U(3)_q\times U(3)_u\times U(3)_d$, which  implies that the explicit PQ violating operators
\be
{\cal O}_{\sslash{\textrm{PQ}}} &=&\det \lambda_u(\phi/m_{\rm Pl}), \quad \det \lambda_d(\phi/m_{\rm Pl})
\ee
can be consistently added to the Lagrangian.  

The dimension of ${\cal O}_{\sslash{\textrm{PQ}}}$ depends strongly on the physics which generates $\lambda_{u,d}$.  For instance, given the ansatz of minimal flavor violation (MFV) \cite{D'Ambrosio:2002ex}, $\lambda_{u,d}$ are linear in the fields $\phi_{u,d}$, whose vevs are the SM Yukawa matrices.  In this case, ${\cal O}_{\sslash{\textrm{PQ}}}$ are dimension 3 operators and thus relevant!  Considering that we need to forbid all explicit PQ violating operators up to dimension $n=10$, this is a complete catastrophe.  

Nevertheless, all is not lost---indeed, the entire enterprise of flavor model-building has been essentially directed at explaining the smallness of the SM Yukawa couplings using non-generic structures that invariably increase the operator dimension of $\lambda_{u,d}$.  As we will see in section \ref{sec:model}, radiative fermion mass generation is an elegant framework in which PQ symmetry protection is straightforwardly accommodated and enforced up to a very high operator dimension.  

\section{Axion Flavor Protection} 
\label{sec:model}

In this section we show how axion flavor protection can be employed to suppress explicit PQ symmetry violating contributions.  Our construction requires two essential ingredients:
\begin{itemize}
\item
{\bf gauged quark flavor symmetry}, under which the PQ  breaking fields are charged.  This will ensure that all explicit PQ violating operators involve all three quark generations.
\item
{\bf radiative quark flavor generation}, which will guarantee that any operator involving the light quark generations has a significantly boosted mass dimension.
\end{itemize}
In what follows, we present a simple description of radiative fermion mass generation from an effective field theory approach.  We then discuss how gauging the quark flavor symmetry can help protect the PQ symmetry.





\subsection{A Low Energy Description of Radiative Flavor}

\label{sec:radeft}

Radiative fermion mass generation is a very old idea \cite{Weinberg:1972ws,Georgi:1972mc,Georgi:1972hy} which explains the hierarchies of the SM Yukawa couplings by relative loop factors.  The heavier fermion generations typically acquire masses at tree-level, while the lighter fermion generations do so via radiative corrections.  To this end, a great deal of work has been devoted to making these constructions realistic and consistent with supersymmetry \cite{Babu:1989tv,ArkaniHamed:1996zw,ArkaniHamed:1995fq,Nandi:2008zw} and grand unification \cite{Barr:2007ma,Barr:2008kg,Ibanez:1981nw}.  Despite the diversity of models, however, they can all be conveniently understood from a simple low-energy perspective.  To begin, let us define the usual quark flavor symmetry of the SM
\be
G_{\rm flavor} &=& SU(3)_q \times SU(3)_u \times SU(3)_d,
\ee
which is a symmetry only in the absence of Yukawa interactions.  Under $G_{\rm flavor}$, the SM fermions and Yukawa matrices have the following transformation properties
\be
\begin{tabular}{lll}
$q = ({\bf 3}, 1,1)$ && $\lambda_u = ({\bf \bar 3},{\bf 3},1)$\\
$\bar u = (1, {\bf \bar 3},1)$ && $\lambda_d = ({\bf \bar 3},1,{\bf 3})$\\
$\bar d = (1, 1,{\bf \bar 3})$&&
\end{tabular}
\ee
where we have defined the spurionic transformation properties of the Yukawa matrices in the spirit of MFV.  

The premise of radiative fermion mass generation is that $\lambda_{u}$ and $\lambda_{d}$ are not the primordial spurions of flavor symmetry breaking.  In fact, they are composed of sums and products of a set of more basic spurions which transform as
\be
\begin{tabular}{lll}
$\phi_q = ({\bf 3}, 1,1)$ && $\alpha_q = ({\bf 8},1,1)$\\
$\phi_u = (1, {\bf 3},1)$ && $\alpha_u = (1,{\bf 8},1)$\\
$\phi_d = (1, 1,{\bf 3})$ && $\alpha_d = (1,1,{\bf 8})$\\
\end{tabular}
\ee
under $G_{\rm flavor}$ and are singlets under SM gauge group.  In other words, there is a $\phi$undamental and $\alpha$djoint spurion field corresponding to each $SU(3)$ factor of $G_{\rm flavor}$.  As we will see shortly, the $\phi$ spurions will ultimately generate the heavy quark masses while the $\alpha$ spurions will induce masses and mixings for the lighter generations.

While in principle $\alpha_{q,u,d}$ can be a complex octet, we choose it to be a real octet for simplicity.  Naively, one might worry whether a real octet is capable of generating the weak CP phase which is known to exist in the CKM matrix.  Nonetheless, despite the nomenclature, a real octet is still a traceless Hermitian $3\times 3$ matrix and thus contains an imaginary component which can successfully produce the measured CKM phase \cite{Masiero:1998yi}.  

One may be slightly discomforted by the sheer number of flavor breaking spurions that we have introduced.  Nonetheless, we have done so only for the sake of generality, and in specific UV completions most of these spurions will actually be equal.  For instance, in section \ref{sec:gaugeflav} we ultimately consider a theory in which $\alpha_{q}=\alpha_{u}^*=\alpha_{d}^*$.

For notational convenience, let us introduce a bra-ket notation in which (bras) kets denote fields transforming in the (anti-)fundamental representation of flavor, so for instance $(\phi_q)_i \equiv |\phi_q \r$ and $(\phi_q^*)^{i}\equiv \l \phi_q |$.  In this notation, the SM Yukawa couplings become
\be
{\cal L} &\supset& \l \bar u | \lambda_u |   q \r h_u +\l \bar d| \lambda_d |  q \r h_d,
\ee
where $SU(2)_L$ indices have been contracted in the obvious way.  

Finally, we have the necessary machinery to construct $\lambda_{u,d}$ from the primordial spurions  $\phi_{q,u,d}$ and $\alpha_{q,u,d}$.  As we will see, the flavor symmetries will completely fix the structure of $\lambda_{u,d}$.  To organize the calculation, let us expand the Yukawa matrices in powers of the $\alpha$ spurions, so that
\be
\lambda_{u,d} &=& \sum_{l=0} \lambda^{(l)}_{u,d},
\ee
where $\lambda^{(l)}_{u,d}$ is $l$th order in the $\alpha$ spurions.  Since $\lambda_u$ transforms as $({\bf \bar 3},{\bf 3},1)$ under $G_{\rm flavor}$, its 0th order component is a bi-triplet constructed from $\phi_q^\dagger$ and $\phi_u$.  Specifically, $\lambda_u^{(0)}$ takes the form
\be \lambda_u^{(0)} &\sim& |\phi_u \r \l \phi_q| \sim 
\left(
\begin{array}{ccc}
0 & 0 & 0 \\
0 & 0 & 0 \\
0 & 0 & 1 \\
\end{array}
\right),
\ee
where we have applied an $SU(3)_q\times SU(3)_u$ in order to rotate $\phi_{q}$ and $\phi_u$ to point in the 3 direction.  This direction in flavor space defines the top quark.  Furthermore, at this order the charm and up quarks are massless.  Contributions at higher order in the $\alpha$ spurions will be generated by radiative corrections and take the form
\be \lambda_u^{(1)} &\sim& \alpha_u|\phi_u \r \l \phi_q| +|\phi_u \r \l \phi_q|\alpha_q\sim 
\left(
\begin{array}{ccc}
0 & 0 & 0 \\
0 & 0 & \alpha \\
0 & \alpha & 0 \\
\end{array}\right)\\
\lambda_u^{(2)} &\sim& \alpha_u^2|\phi_u \r \l \phi_q| + \alpha_u|\phi_u \r \l \phi_q|\alpha_q+ |\phi_u \r \l \phi_q|\alpha_q^2\sim 
\left(
\begin{array}{ccc}
0 & 0 & \alpha^2 \\
0 & \alpha^2 & 0\\
\alpha^2 & 0 & 0 \\
\end{array}
\right) \\
&\vdots &
\ee
and so on and so forth.  As before, we have gone to a convenient basis at each step in order to define new flavor directions corresponding to the charm and up quark.  In particular, after $\lambda^{(0)}_u$ is generated, there is still a residual $SU(2)_q\times SU(2)_u$ symmetry which corresponds to the un-lifted degeneracy between the charm and up quarks.  This residual symmetry can be used to rotate $\lambda^{(1)}_u$ into the canonical form shown above.  Thus, $\lambda_u^{(1)}$ contributes to the top-charm mixing angle, while $\lambda_u^{(2)}$ contributes to the top-up mixing angle and the charm mass.  Summing contributions at all orders in the $\alpha$ spurions, we find
\be
\lambda_{u} \sim \left(
\begin{array}{ccc}
\alpha^4 & \alpha^3 & \alpha^2 \\
\alpha^3 & \alpha^2 & \alpha\\
\alpha^2 & \alpha & 1 \\
\end{array}
\right),
\ee
with an analogous construction for $\lambda_d$.  Now in reality, each insertion of $\alpha$ in the above expressions is accompanied by its own order one coefficient, in addition to $(4\pi)^2$ factors and logarithms arising from actual loop diagrams.  Since these factors are sensitive to the particulars of the UV completion, we suppress these various factors for simplicity.  Nevertheless, $\alpha$ can be thought of as a small parameter of flavor breaking, and thus we have  successfully generated the hierarchy of SM quark masses and mixings!

\subsection{Gauging the Flavor Symmetry}
\label{sec:gaugeflav}

Up till now we have made no assumptions about the UV beyond those required for radiative fermion mass generation---this will now change.  In particular, axion flavor protection requires an additional but essential ingredient, which is a gauged flavor symmetry.  Consider the following gauge group
\be  
G_{\rm gauge} &=& SU(3)_f \times U(1)_X \\
 SU(3)_f &\equiv &\textrm{diag}[ \{ SU(3)_q,SU(3)_u^*, SU(3)_d^*\}],
\ee
where $SU(3)_f$ is the diagonal subgroup of the $SU(3)^3$ flavor symmetry\footnote{While gauging the quark flavor symmetry does not induce a gauge anomaly for $SU(3)_c$, will must still add spectator fermions to cancel the gauge anomaly for $G_{\rm flavor}$.}.  As we will see later, this choice is natural in grand unified theories \cite{Grinstein:2006cg} since $q$ and $\bar u$ live in the same multiplet.   Under $G_{\rm gauge}$, the SM fermions all transform in the fundamental representation of $SU(3)_f$ and are $U(1)_X$ singlets:
\be
\begin{tabular}{l}
$q = ({\bf 3}, 0)$ \\
$\bar u = ( {\bf 3},0)$ \\
$\bar d = ({\bf 3},0)$
\end{tabular}
\ee
Furthermore, let us introduce the following scalar fields
\be
\begin{tabular}{lll}
$\phi_+ = ({\bf 3}, 1/6)$ && $\alpha = ({\bf 8},0)$\\
$\phi_- = ({\bf 3},-1/6)$ && $\chi = (1,2)$
\end{tabular}
\ee
which acquire vevs due to a Mexican hat potential.  We will assume that the vevs of $\phi_+$ and $\phi_-$ are aligned, which can be easily accommodated given the appropriate potential.  The vevs of $\phi_\pm$ and $\alpha$ will ultimately generate the Yukawa matrices of the SM, as described in section \ref{sec:radeft}.

Spontaneous symmetry breaking implies the existence of Nambu-Goldstone bosons---some will be eaten due to the Higgs mechanism and others will remain in the spectrum as physical degrees of freedom.  For the purposes of our discussion, we will only be concerned with three particular modes:
\be
\phi_\pm = 
\left(
\begin{array}{c}
0 \\
0 \\
f_\pm \\
\end{array}
\right)e^{i a_\pm/f_\pm} ,\qquad \chi = f'  e^{i a' /f'}.
\ee
Via the Higg mechanism, two linear combinations of $a_+$, $a_-$, and $a'$ are eaten by the $T_8$ gauge boson of $SU(3)_f$ and the gauge boson of $U(1)_X$.  The remaining physical degree of freedom is the axion, $a$, and it resides in all three primordial fields---that is to say, the overall phases of $\phi_+$, $\phi_-$, and $\chi$ each contain some component of the axion.  The relative admixture  will be unimportant for the following discussion.  Also, note that it was necessary to include the flavor singlet $\chi$ so that a physical axion remained in the spectrum.

\subsection{The Leading Dangerous Operator}

Let us now construct the lowest dimension gauge invariant operator which induces a mass for the axion.  Such an operator will depend on $\phi_\pm$ and $\chi$ and will necessarily violate the PQ symmetry.  Since the axion resides in the phases of $\phi_\pm$ and $\chi$, it is obvious that gauge invariant quantities such as $\phi_\pm^\dagger \phi_\pm$ and $|\chi|^2$ are harmless since they have no axion dependence whatsoever.  On the other hand, operators such as $\phi_\pm^\dagger \phi_\mp$ depend on the axion but are not gauge invariant, since they have non-zero $U(1)_X$ charge.  Stringing together operators of this type, however, one can construct a gauge invariant operator, $(\phi^\dagger_+ \phi_-)^6 \chi$, which is dimension $n=13$ and thus sufficiently small to preserve the axion solution to the strong CP problem.  The high dimension of this operator is a consequence of the gauged $SU(3)_f$ and our choice of $U(1)_X$ charges.

However, our work is not done---there exist additional gauge invariant operators involving the epsilon tensor of $SU(3)_f$ which can also potentially destabilize the axion potential.  At leading order in $\phi_\pm$, one can write down the simplest $SU(3)_f$ invariant of this form
\be
\epsilon^{ijk} \phi_{+i} \phi_{+j} \phi_{+k} = \det (\, |\phi_+\r\, |\phi_+ \r\, |  \phi_+ \r\,)=0.
\ee
Fortunately, this quantity is  zero, since it is the determinant of a matrix whose columns are linearly dependent.  Note that this is still the case if we replace any $+$ with $-$, since $\phi_+$ and $\phi_-$ have aligned vevs.
\begin{figure}[t]
\begin{center}
 \includegraphics[scale=0.8]{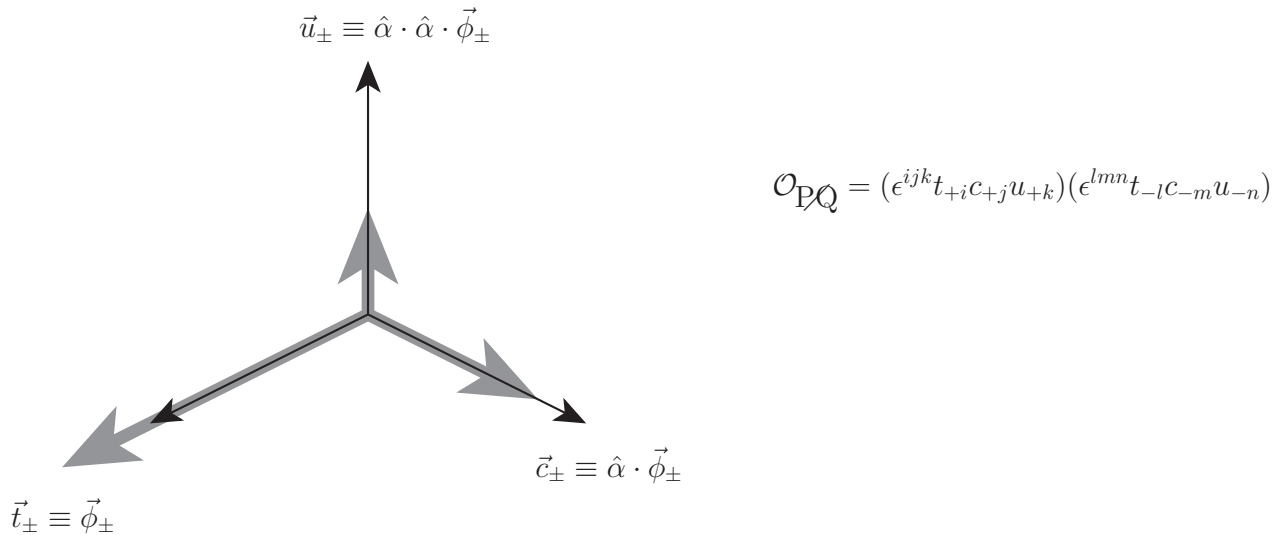} 
\end{center} 
\caption{
Radiative corrections generate three linearly independent vectors in flavor space which define the top, charm, and up quark.  Due to $G_{\rm gauge}$,  the leading gauge invariant, explicit PQ violating operator must involve all three generations and is dimension $n=12$.} 
\end{figure}
Clearly, any non-vanishing gauge invariant operator involving $\epsilon^{ijk}$ will have to involve three linearly independent vectors in flavor space, which is just to say that it needs to know about the top, charm, and up quark!  For this reason, the leading gauge invariant, explicit PQ violating operator is of the form\footnote{There also exist explicit PQ violating operators which involve derivatives, but these do not contribute to the axion potential beyond those operators which we consider here (see appendix \ref{app2}).}
 \be 
{\cal O}_{\sslash{\textrm{PQ}}} &=& \det(\, |\phi_+ \r\, |\alpha \phi_+ \r\, |\alpha^2 \phi_+ \r\,) \times \det(\, |\phi_- \r\, |\alpha \phi_- \r\, |\alpha^2 \phi_- \r\,), 
\ee
where any $+$ and $-$ can be swapped to yield an equally valid gauge invariant operator.  Since this operator is dimension $n=12$, the axion mass is sufficiently protected.

Notice how ${\cal O}_{\sslash{\textrm{PQ}}}$ is {\it precisely} of the form of $\det M \sim \det \lambda_u \times \det \lambda_d$.  Hence, the {\it very same} flavor structures which ensure naturally small Yukawas also contribute to the extremely high mass dimension of the leading explicit PQ violating operator.

\subsection{A Simple Renormalizable UV Completion}
\label{sec4}

Let us now turn to a simple renormalizable UV completion.  Since the bulk of literature concerning radiative fermion mass generation simply assumes the existence of a primordial top Yukawa, this UV completion can be easily ``retro-fit'' onto any number of the existing theories.

In particular, we can retrofit the Yukawa interactions for the top and bottom quarks with the only allowed gauge invariant dimension 6 operators
\be \lambda_{t} \, q_3 h_u \bar u_3 &\rightarrow& \l \phi_\mp | \bar u \r \l \phi_\pm | q  \r h_u\\
\lambda_{b} \, q_3 h_d \bar d_3 &\rightarrow& \l \phi_\mp | \bar d \r \l \phi_\pm | q  \r h_d,
\ee
where (bras) kets represent (anti-)fundamentals of $SU(3)_f$.
\
The vev of $\phi_\pm$ generates the heavy quark Yukawa couplings.  Since $\alpha$ also couples to the quarks via $SU(3)_f$ gauge interactions, the Yukawa couplings for the light fermion generations will be induced radiatively, yielding the hierarchical pattern for $\lambda_u$ and $\lambda_d$ discussed in section \ref{sec:radeft}.  Because the focus of present work is PQ symmetry protection rather than flavor generation, we will not concern ourselves with the details of engineering a completely realistic CKM matrix.

The above dimension 6 operators will arise from the following renormalizable Lagrangian
\be
&{\cal L}_{\rm UV} \supset \l \phi_\mp | \bar u\r U' + \bar U' \l  \phi_\pm | U \r + \l \bar U | q\r h_u 
- M\l \bar U| U\r - M' \bar U' U'& \\
&\Downarrow& \\
&{\cal L}_{\rm IR} \supset  \frac{1}{M M'} \l \phi_\mp | \bar u \r \l \phi_\pm | q  \r h_u, &
\ee
where $U$ and $U'$ are new fermions which have the electroweak quantum numbers of the up quark, and $U'$ is also $U(1)_X$ charged.  Of course, in order for $\lambda_t\sim 1$, it must be the case that $f_+ f_-/(MM')\sim1$. An analogous construction can of course be employed for the bottom Yukawa.  Thus, we see that the heavy quark Yukawa couplings are simultaneously generated by the PQ symmetry breaking.

This story is naturally extended to theories with grand unification.  For example, in $SU(5)$ GUT it is natural to gauge $SU(3)_f$ simply because $q$ and $\bar u$ reside in the same 10 of $SU(5)$ and thus have identical flavor quantum numbers. In this instance, we can retrofit the $SU(5)$ Yukawa couplings as in the previous example,
\be
\lambda_t \, 10_3 \, 5\, 10_3 &\rightarrow& \l \phi_\mp|10 \r \l\phi_\pm | 10 \r 5 \\
\lambda_{b/\tau} \, 10_3 \, \bar 5 \, \bar 5_3 &\rightarrow& \l \phi_\mp| \bar 5 \r \l\phi_\pm | 10 \r \bar 5 ,
\ee
where 5, $\bar 5$, and 10 label representations under $SU(5)$ and the 3 subscripts denote the third generation.  

\section{Discussion}
\label{sec:conclusion}

The efficacy of the QCD axion is highly sensitive to the quality of the PQ symmetry.  We have identified an irreducible source of explicit PQ violation which is present in all theories and must be adequately suppressed in order to solve the strong CP problem.  Furthermore, we have proposed a mechanism whereby explicit PQ violating operators are forbidden with the help of the same flavor structures which generate the hierarchy among the SM quark masses and mixings.  

A universal attribute of this construction is that the Yukawa couplings of the heavy quark generations are induced at the PQ breaking scale.  For this reason, this setup can be retro-fit onto existing models of radiative fermion mass generation in which the heavy quark Yukawas are input by hand.

While this mechanism is simply incorporated into theories of grand unification, extensions to SUSY theories are more limited because of the intrinsic tension between SUSY and radiative fermion mass generation \cite{ArkaniHamed:1996zw,ArkaniHamed:1995fq}.  In particular, since SUSY renormalization theorems imply that the SM Yukawas are impervious to radiative corrections in the SUSY limit, any relevant effect must be proportional to SUSY breaking.  Due to severe experimental constraints on squark-induced FCNCs, any phenomenologically viable SUSY version of our setup would necessarily involve a split spectrum \cite{ArkaniHamed:2004yi,ArkaniHamed:1999za}.

Finally, extending this construction to FN theories and extra-dimensional models of flavor is another possibility which is beyond the scope of this work, but may require further study.

\section*{Acknowledgements}

It is a pleasure to thank M. Dine and K. S. Babu for helpful discussions and comments on the draft.  We especially thank J. Thaler for many helpful discussions and comments, as well as collaboration at the early stages of this work. The work of C.C. was supported by the National Science Foundation under grant PHY-0555661.

\appendix

\section{Derivative Operators and Explicit PQ Violation}

\label{app2}

The attentive reader will notice that we have ignored the presence of explicit PQ violating operators which involve derivatives.  In this appendix we show that such operators exist but do not destabilize the axion potential.

The contributions we are concerned with are derivatively coupled, explicit PQ violating operators of the form
\be
{\cal O}_{\sslash{\textrm{PQ}}} &=& S_{\mu\nu\rho}S^{\mu\nu\rho} \\
S_{\mu\nu\rho} &=& \det(\, | \phi \r\, |\partial_\mu\phi \r\, |\partial_\nu \partial_\rho\phi \r\,),\ee
where we have suppressed $+$ and $-$ subscripts.  Can operators of this type displace the axion potential?  For example, if ${\cal O}_{\sslash{\textrm{PQ}}}$ is added to the Lagrangian, then an axion potential is naively generated via Feynman diagrams involving loops of heavy field fluctuations and this operator.  

Nevertheless, there is a simple reason for why these radiative corrections do not introduce any operators we have not already considered.  In particular, the real question is whether this operator will, via RG flow, induce a potential for the {\it zero modes} of $\phi$, i$.$e$.$ a CW potential.  From this point of view we are simply integrating out momentum shells in the Wilsonian sense, and then restricting to the zero modes of all fields.  However, we have already catalogued all the gauge invariant explicit PQ violating operators which can appear in the CW potential---and they are all have mass dimension $n=12$ or greater.

\bibliographystyle{prsty}
\bibliography{flavor_cp_bib}
\end{document}